\documentclass[12pt, letterpaper, ]{article}

\usepackage[utf8]{inputenc}
\usepackage{amssymb}
\usepackage{multicol}
\usepackage[margin=2.35cm]{geometry}
\usepackage{setspace}
\usepackage{graphicx}
\usepackage{url}
\usepackage{color}
\usepackage{hyperref}
\graphicspath{ {images/} }
\usepackage{tikz}
\usepackage{amsmath}
\usepackage{physics}
\usepackage{ulem}

\begin{document}

\title{Einstein Completeness as Categoricity}

\author{Iulian D. Toader}

\date{\small Institute Vienna Circle, University of Vienna \\ iulian.danut.toader@univie.ac.at}

\maketitle

\begin{abstract}

This paper provides an algebraic reconstruction of Einstein's own argument for the incompleteness of quantum mechanics -- the one that he thought did not make it into the EPR paper -- in order to clarify the assumptions that underlie an understanding of Einstein completeness as categoricity, the sense in which it is a type of descriptive completeness, and some of the various ways in which it has been more often misconstrued.
\end{abstract}

\doublespacing

\section{Introduction}

Einstein's argument for the incompleteness of quantum mechanics, which did not make it into the EPR paper (Einstein, Podolski, and Rosen 1935) in the way Einstein thought it would, was clearly formulated in letters to Schrödinger and Popper, as well as in several publications (e.g. Einstein 1936). After Arthur Fine brought it to philosophical attention (Fine 1981), Don Howard suggested that the argument might be understood as deploying a notion of completeness known as categoricity, i.e., model uniqueness up to isomorphism (Howard 1990). This suggestion was motivated by Einstein's claim that quantum mechanics fails to assign a unique wavefunction to the real state of one subsystem of an EPR system, since the assignment depends on the measurement that could be performed on the other subsystem. If multiple wavefunctions can be assigned to the same subsystem, and if one is justified in considering them as non-isomorphic models, then this would be enough to show quantum mechanics non-categorical. If Howard's suggestion is taken seriously, then Einstein completeness turns out to be a rather different type of completeness than the one articulated in the EPR paper. 

In the present paper, I provide an algebraic reconstruction of Einstein's argument, which I think can clarify the assumptions underlying an understanding of Einstein completeness
as categoricity. The key idea to be rigorously articulated is the following: ``if one understands a theoretical state as, in effect, a model for a set of equations plus boundary conditions ..., then Einstein's conception of a completeness requirement should really be understood as a categoricity requirement.'' (Howard 1992, 208) To stay as close as possible to Einstein's own argument, I focus on the original EPR state, with observables having a continuous spectrum, suitably defined within an algebraic framework (Arens and Varadarajan 2000, Werner 1999), and I explain under what conditions one would be justified to read Einstein completeness as categoricity. On my reconstruction, the
argument assumes that representations on (tensor products of) Hilbert spaces are the models of quantum mechanics of (composite) systems. It assumes as well that the unitary equivalence of such representations is a necessary (though not a sufficient) condition for categoricity. The argument then points out that there are representations of the (tensor product of) algebras describing a subsystem of an EPR system that are not unitarily equivalent. As I reconstruct it, following Howard's suggestion, Einstein's argument concludes that categoricity is logically inconsistent with separability and locality.

It is, of course, difficult to say that the reconstruction I propose is actually entirely faithful to Einstein's own thought. One worry one might raise is that while my reconstruction works for the original EPR state, it does not work for entangled spins, since on finite-dimensional Hilbert spaces there are no unitarily inequivalent representations. But Einstein, as is well known, never cared much about Bohm's version of the EPR argument.\footnote{The only place where Einstein formulated a spin version of his argument appears to be in a late manuscript from around 1955: for discussion, see Sauer 2007.} Focusing on the infinite-dimensional case is thus historically reasonable. Moreover, for Einstein's objection to stand, it is of course sufficient that his argument goes through in one case; it is not required that it should do so in all cases. In any event, I will argue that my reconstruction is preferable to others, according to which Einstein's argument should be taken to establish  ``overcompleteness'' (Lehner 2014) or unsoundness (Gömöri and Hofer-Szabó 2021), rather than non-categoricity.  Furthermore, I suggest that my reconstruction sheds some new light on the Bohr-Einstein controversy. Bohr's doctrine of complementarity has been interpreted in terms of the unitary inequivalence of non-regular Hilbert space representations (Halvorson 2004), an interpretation that vindicates a common view that Bohr's notion of completeness was significantly distinct from the descriptive completeness articulated in the EPR paper.\footnote{For an expression of this view, see, e.g., Norsen 2017, 148.} But on my reconstruction, Einstein completeness fails precisely due to this unitary inequivalence. Thus, from an algebraic point of view, it appears that the sense in which Bohr thought quantum mechanics was complete is exactly the sense in which Einstein argued it wasn't. Considered from a model-theoretical perspective, their views would turn out to be precisely antithetical. Of course, in order to properly justify this claim, a more formal undertaking would be needed than what this paper can offer. In particular, one would need a fully articulated formal semantics of standard quantum mechanics that deploys a suitable logic for a formalization of the algebraic structures described below. Deferring this undertaking to further work seems, however, in order for now.

\section{Einstein completeness}

Recall the completeness condition that the EPR paper purported to argue it is not satisfied by quantum mechanics: ``Whatever the meaning assigned to the term \textit{complete}, the following requirement for a complete theory seems to be a necessary one: \textit{every element of the physical reality must have a counterpart in the physical theory}.'' (1935, 777) This condition has typically been understood to formulate a type of descriptive completeness, since it applies to a physical theory just in case that theory is able to describe all of the physical reality that it aims to describe. Related ideals of completeness were endorsed by other foundational thinkers, before and after the 1930s, within and without quantum physics, and particularly with respect to systems of axioms. For instance, Russell and Whitehead stated that a system of axioms for pure mathematics (like that presented in their \textit{Principia Mathematica}) is complete just in case it is able to capture ``as much as may seem necessary'' of the domain that it aims to describe, i.e., the entire class of theorems of ordinary mathematics. This was a notion of completeness that G\"{o}del took to prove that, on some reasonable assumptions, it cannot be attributed to the system in the \textit{Principia}.\footnote{See (Detlefsen 2014) for a discussion of completeness as understood by Russell and Whitehead, and an argument that G\"{o}del's own understanding of completeness, as negation-completeness, is relevantly different than the descriptive variety.} Similarly, von Neumann considered a system of axioms for quantum mechanics (like that presented in his \textit{Mathematical Foundations of Quantum Mechanics}), to be complete just in case it is able to derive all statistical formulas of quantum mechanics, and rejected the view that their statistical nature was due to ``an ambiguity (i.e., incompleteness) in our description of nature''.\footnote{See (Acuña 2021) for a recent re-evaluation of (the debate on) von Neumann's proof of completeness.} 

The EPR paper argued that quantum mechanics does not satisfy the completeness condition because in the case of a system in an EPR state there are elements of physical reality, i.e., properties of a subsystem of that system, that the theory aims to describe, but fails to do so. As is well known, crucial to the EPR argument is the following criterion of reality: ``If, without in any way disturbing a system, we can predict with certainty (i.e., with probability equal to unity) the value of a physical quantity, then there exists an element of physical reality corresponding to this physical quantity.'' (1935, 777; italics removed) The role and character of this criterion, as well as the formal structure of the argument, have been often addressed.\footnote{Perhaps the most detailed formal reconstruction of the EPR argument has been given in (McGrath 1978) where the following is also noted: ``Regrettably EPR equate two notions of completeness: `complete representation by a
wave function' and `complete theory' are used interchangeably.'' (560) See also (Gömöri and Hofer-Szabó 2021) for a nice discussion of the EPR criterion of reality, its indispensable role within the EPR argument for incompleteness, and its absence from Einstein’s own argument.} What has been often ignored is the fact that EPR completeness and Einstein completeness are not the same type of \textit{descriptive} completeness.\footnote{The fact that EPR completeness and Einstein completeness are not the same type of completeness has been, to my knowledge, first explicitly noted by Arthur Fine:  ``Einstein does not give [the latter] a catchy name, but ... [we can] call this more technical conception \textit{bijective completeness}.'' (Fine 1981, 72)} In one widely quoted passage from his letter to Schrödinger, Einstein writes: ``In the quantum theory, one describes a real state of a system through a normalized function, $\psi$, of the coordinates (of the configuration-space). ... Now one would like to say the following: $\psi$ is correlated one-to-one with the real state of the real system. ... If this works, then I speak of a complete description of reality by the theory. But if such an interpretation is not feasible, I call the theoretical description `incomplete'.'' (Letter to Schrödinger, 19 June 1935; translated in Howard 1985, 179) Einstein completeness can thus be attributed to quantum mechanics if and only if there exists a one-to-one correlation between a $\psi$-function and the real state of a system the theory aims to describe. Einstein's point that a one-to-one correlation does not exist in the case of an EPR system is meant to be supported by his separability and locality assumptions, so his argument can be formulated in the following way:

\bigskip

1. Spacelike separated physical systems have real states, which cannot causally influence one another.

2. Consider a system with two subsystems, A and B, in an EPR state.

3. Thus, each subsystem has a real state, no matter what measurements can be carried out on its other subsystem.

4. Quantum mechanics assigns different $\psi$-functions to A, depending on which observable one can choose to measure on B.

5. But if quantum mechanics is complete, these $\psi$-functions should be identical.

6. Thus, quantum mechanics is incomplete.

\bigskip

Most of my discussion in this paper will be focused on premises 4 and 5. It should be clear that the notion of completeness in premise 5 is directly implied by that defined in the letter to Schrödinger. In a letter to Pauli, Schrödinger put it in the following metaphorical terms: ``He [i.e., Einstein] has a model of that which is real consisting of a map with little flags. To every real thing there must correspond on the map a little flag, and vice versa.'' (Schrödinger to Pauli, July 1935, translated in Howard 1990, 106) It immediately follows that a map with little flags is an Einstein complete description of physical reality only if the map is essentially unique, in the sense that there is a bijection between any two maps describing that reality. In the case of quantum mechanics, by analogy, Einstein completeness requires that the $\psi$-function representing the real state of a system be essentially unique.

\newpage

Howard's suggestion as to how to properly understand Einstein completeness, first offered in a footnote to his presentation of Einstein's argument, is this: ``A complete theory assigns one and only one theoretical state to each real state of a physical system. [Footnote: This is a curious conception of completeness, more akin to what is called in formal semantics ``categoricity.''] But in EPR-type experiments involving spatio-temporally separated, but previously interacting systems, A and B, quantum mechanics assigns different theoretical states, different `psi-functions,' to one and the same real state of A, say, depending upon the kind of measurement we choose to carry out on B. Hence quantum mechanics is incomplete.'' (Howard 1990, 64) The same suggestion is then uplifted from footnote to the main text of a later paper: ``Let me conclude with one really outrageous suggestion. ... the operative criterion of completeness in Einstein's thinking was this: a theory is complete if and only if it assigns a unique theoretical state, such as a psi-function, to every unique real physical state. But if one understands a theoretical state as, in effect, a model for a set of equations plus boundary conditions ..., then \textit{Einstein's conception of a completeness requirement should really be understood as a categoricity requirement}. In other words, Einstein is saying that a `complete' (read `categorical') theory is one that determines a unique (\textit{eindeutige}) model for the reality it aims to represent.'' (Howard 1992, 208; my emphasis) The conception of completeness that Howard thought should be credited to Einstein is that according to which a theory is Einstein complete if and only if all its models belong to one and the same isomorphism class. 

In the background to Howard's suggestion lies, of course, a complex model-theoretical machinery. Models are defined as set-theoretical structures of elements in a domain that can be assigned to a formal version of the language of quantum mechanics (including symbols for variables, constants, functions, and relations) by an interpretation map. But in order to reach a good understanding of Howard's suggestion, starting from my reconstruction of Einstein's argument, we will not actually need this formal machinery. It will be enough to motivate informally an algebraic requirement for categoricity, and then argue that Einstein can be understood to have doubted that quantum mechanics satisfies that requirement. It will then follow that what Einstein argued for is that quantum mechanics is descriptively incomplete in the sense that, roughly, even though it may be able to describe all of the physical reality that it aims to describe, it ends up describing a lot more besides that as well.

The reconstruction of Einstein's argument, in an algebraic framework, will be given further below. That will require a preliminary consideration of premise 2, in particular, a properly algebraic definition of the original EPR state, for continuous observables. It will also require a discussion of premise 4, that is an algebraic account of the difference between the $\psi$-functions assigned to susbsystem A. The outcome of all that will, I hope, be three-fold:

\bigskip

(i) a good understanding of the assumptions behind Howard's suggestion to read Einstein completeness as categoricity, 

(ii) an explanation of what makes Einstein completeness different from EPR completeness as a type of descriptive completeness, and finally, 

(iii) an unequivocal sense of where some of the misreadings of Einstein completeness have gone wrong. 

\bigskip

Although my real emphasis will be on (i), let me briefly address (iii) right away, in order to candidly raise the reader's interest in what is to come later, in section 4, which takes care of (i). Then (ii) will be presented in section 5, where I also briefly indicate how (i) bears on the Bohr-Einstein debate.

\section{Misconstruals}

Einstein's argument for incompleteness is sometimes considered too confused to take seriously. Klaas Landsman, for example, maintained that it is a ``muddled'' argument (Landsman 2006, 234) for the following reason: ``Unfortunately, Einstein (and EPR) insisted on a further elaboration of this disjunction [i.e., completeness or separability], namely the idea that there exists some version of quantum mechanics that is separable ... at the cost of assigning more than one state to a system (two in the simplest case). It is this unholy version of quantum mechanics that Einstein (and EPR) called `incomplete'. Now, within the formalism of quantum mechanics such a multiple assignment of states (except in the trivial sense of wave functions differing by a phase factor) makes no sense at all, for the entanglement property lying at the root of the non-separability of quantum mechanics is so deeply entrenched in its formalism that it simply cannot be separated from it.'' (\textit{op. cit.}, 227) However, if it implies that Einstein's argument requires that entanglement be relinquished, this is simply missing the point. One would have thought it obvious that, quite the contrary, the argument is essentially based on entanglement. Clearly, Landsman's criticism fails insofar as it neglects the fact that Einstein separability (expressed in premise 1 above) and entanglement (implicit in premise 2) are compatible.\footnote{See (Murgueitio Ram\'irez 2020) for the point that Einstein's argument would be trivially unsound, if separability and entanglement were not compatible. Landsman argues that, as a consequence of Raggio's theorem, Einstein separability is mathematically equivalent to Bohr's doctrine of the necessity of classical concepts. But his argument takes Einstein separability as state decomposability.} The alleged reason why multiple assignment of states is nonsensical doesn't stand: there is no ``unholy version'' of quantum mechanics, i.e., a version of the theory without entanglement, and Einstein did not think otherwise. Furthermore, and this is a more significant problem, Landsman seems to misconstrue Einstein's view on the difference between the $\psi$-functions assigned to subsystem A as a trivial phase difference. As we will see further below, the multiple assignment of $\psi$-functions that Einstein actually considered can be aptly interpreted algebraically in terms of the unitary inequivalence of non-regular Hilbert space representations. Were this nonsensical, Bohr's complementarity doctrine (at least as interpreted in Halvorson 2004, see below section 5) would appear to be nonsensical as well, but this is a doctrine that Landsman professes to defend.

In \textit{The Cambridge Companion to Einstein}, Christoph Lehner characterized Einstein completeness, and Einstein's argument that this cannot be attributed to quantum mechanics, in these terms: ``Einstein ... concludes that the quantum mechanical description is not biunique because it is incomplete. ... This conclusion is not warranted logically.'' (Lehner 2014, 334) A description that is not biunique is one that assigns to subsystem A a $\psi$-function which is not correlated one-to-one with its real state. The bottom half of Einstein's argument would, on this reconstruction, be as follows:

\bigskip

4$_{L}$. Quantum mechanics assigns different $\psi$-functions to A, depending on which observable one can choose to measure on B.

5$_{L}$. Thus, quantum mechanics is incomplete.

6$_{L}$. Thus, quantum mechanics is not biunique.

\bigskip

Lehner maintained that the inference from 5$_{L}$ to 6$_{L}$ is not justified. But this clearly puts Einstein's cart before his horse: the one-to-one correlation does not fail because of incompleteness; rather, its failure is the actual reason for incompleteness, just as the reason why the one-to-one correlation fails is provided by the possible assignment of different $\psi$-functions to A (expressed by 4$_{L}$). Moreover, since a $\psi$-function is just a theoretical description and such ``a description that is not invariant is not necessarily incomplete ... it is `overcomplete' or nonabsolute'' (\textit{loc. cit.}), i.e., it assigns to A a theoretical state that is correlated many-to-one with its real state, what Einstein should have allegedly argued is the following:

\bigskip

4$^{L}$. Quantum mechanics assigns different theoretical states to A, depending on which observable one can choose to measure on B.

5$^{L}$. Thus, quantum mechanics is overcomplete.

6$^{L}$. Thus, quantum mechanics is not biunique.

\bigskip

The inference from 5$^{L}$ to 6$^{L}$ is immediate, but trivial. However, Lehner appears to reduce Einstein's incompleteness to a quite uninteresting case of empirical underdetermination. Jos Uffink has recently pointed out that reading Einstein incompleteness as overcompleteness is ``surprising'', since ``cases of overcompleteness are ubiquitous in physics... Indeed one might wonder whether overcompleteness
is a worrisome issue at all in theoretical physics.'' (Uffink 2020, 557) In quantum mechanics, as Uffink emphasizes, such cases of overcompleteness as illustrated by phase differences between $\psi$-functions are not worrisome at all. But, of course, Einstein did not think otherwise. As already noted, we will see below that the essential differences between $\psi$-functions that Einstein did think worrisome can be aptly interpreted algebraically in terms of the unitary inequivalence of non-regular Hilbert space representations. If this is enough to conclude that quantum mechanics is not categorical, then Lehner's reconstruction of Einstein's argument turns out to conflate (the problem of) categoricity and (the problem of) empirical underdetermination.  

Along different lines, Márton Gömöri and Gábor Hofer-Szabó argue for a similar conclusion, that Einstein's argument should not really be taken to establish incompleteness: ``According to Einstein's later [than the EPR] argument, the Copenhagen interpretation is committed to the existence of elements of reality that cannot be out there in the world –- under the assumptions of locality and no-conspiracy. Hence, given these assumptions, the Copenhagen interpretation is \textit{unsound} –- as opposed to being incomplete.'' (Gömöri and Hofer-Szabó 2021, 13453) Supposing it right that Einstein intended his argument against the Copenhagen interpretation of quantum mechanics, the bottom half of Einstein's argument would change as follows:

\bigskip

4$_{G}$. Quantum mechanics assigns different $\psi$-functions to A, depending on which observable one can choose to measure on B.

5$_{G}$. Thus, at least some $\psi$-functions represent states of A that cannot exist.

6$_{G}$. Thus, quantum mechanics is unsound.

\bigskip

Gömöri and Hofer-Szabó readily acknowledge that their reconstruction of Einstein's argument is in conflict with his own understanding of completeness. Pursuing ``logical reconstruction'' in spite of what ``Einstein actually thought to argue'', they find his own notion ``not quite apt'', in part because it is not identical to the EPR notion of completeness (\textit{loc. cit.}). But one would have thought that the difference between EPR completeness and Einstein completeness cannot be sufficient to deny the latter as an apt notion, since there is nothing to indicate that EPR completeness is the only possible type of descriptive completeness attributable to physical theories. Still, their main reason for having Einstein's argument conclude that quantum mechanics is unsound, rather than incomplete, is the incompatibility of the different $\psi$-functions assigned to subsystem A of an EPR system. This is correct as an understanding of the essential difference between $\psi$-functions, expressed by premise 4$_{G}$. In the next section, this incompatibility will be precisely interpreted algebraically as the unitary inequivalence of non-regular Hilbert space representations. If my interpretation is adequate, I do not see why that incompatibility would make quantum mechanics unsound, rather than sound but non-categorical. To argue that my interpretation is adequate, I turn now to reconstructing Einstein's argument in the framework of C$^{*}$-algebras. 

\section{An algebraic reconstruction of Einstein's argument}

Taking Howard's suggestion seriously requires, as I pointed out in the introduction, that one consider the theoretical states assigned to a system by quantum mechanics as models of the physical theory. But it's not immediately clear how a $\psi$-function, and in particular one that represents the EPR state of a system, can be considered in this way. For one thing, such a $\psi$-function cannot be construed as a unit vector in a standard infinite-dimensional Hilbert space $\mathcal{H}$. This is for multiple related reasons, such as that the $\psi$-function has infinite norm, and the observables (like position and momentum) of the subsystems of an EPR system have continuous spectra, so their probability distribution will have a probability density that cannot be concentrated on a single point in $\mathcal{H}$. This is why the definition of the EPR state for a system composed of subsystems A and B as a unit vector in $\mathcal{H}_{A} \otimes \mathcal{H}_{B}$ for observables in subalgebras $\textbf{B} (\mathcal{H}_{A}) \otimes \mathcal{I}$ and $\mathcal{I} \otimes \textbf{B}(\mathcal{H}_{B})$ is not entirely adequate (Arens and Varadarajan 2000, 638). 

But that definition has been generalized, as we will presently see, and the EPR state has been identified with a normalized positive linear functional on $\mathfrak{A}_{A} \otimes \mathfrak{A}_{B}$ for observables in the von Neumann algebras $\pi (\mathfrak{A}_{A} \otimes \mathcal{I})''$ and $\pi (\mathcal{I} \otimes \mathfrak{A}_{B})''$ (Werner 1999, Halvorson 2000). The generalized definition will be used below to reformulate premises 4 and 5 in Einstein's argument. The essential difference between the theoretical states assigned to subsystem A of the EPR system will be interpreted in terms of the unitary inequivalence of non-regular representations of $\mathfrak{A}_{A} \otimes \mathcal{I}$ on $\mathcal{H}_{A} \otimes \mathcal{I}$. This can support a model-theoretical approach to Einstein's argument, suggested by Howard, provided one takes representations as models of quantum mechanics. But this would not be at all unusual. Consider, for example, R.I.G. Hughes' view: ``Quantum mechanics, we may say, uses the \textit{models} supplied by Hilbert spaces.'' (Hughes 1989, 79)\footnote{Similarly, but more recently: ``The models of NRQM are Hilbert spaces, along with a suitable subalgebra of the bounded operators on that Hilbert space.'' (Weatherall 2019, 7)} These models are assumed to be definable as formal structures in a suitable metatheory of quantum mechanics, but as already noted at the outset, for my purposes in this paper no such definitions are needed.\footnote{This assumption would be problematic, if one dispensed with Hilbert space representations in the formalism of quantum mechanics. For the $^{*}$-automorphisms of the Weyl algebra are unique up to unitary equivalence, so at the abstract algebraic level there are no non-regular states. But I think this problem for my reconstruction can be ignored, since Einstein, interested as he was in local wavefunctions, was surely not an \textit{Algebraic Imperialist} of the kind introduced and described in Arageorgis 1995 (see also Ruetsche 2011).}

Let's start by considering observables $O_{A}\otimes I \in \textbf{B} (\mathcal{H}_{A}) \otimes \mathcal{I}$ and $ I \otimes O_{B} \in \mathcal{I} \otimes \textbf{B} (\mathcal{H}_{B})$. As defined by Richard Arens and Veeravalli S. Varadarajan, $\omega$ is an EPR state if and only if the joint distribution of $O_{A}\otimes I$ and $ I \otimes O_{B}$ is a measure on $\mathbb{R}^{2}$ concentrated on the diagonal, $\mu_{\omega}^{O_{A}\otimes I, I \otimes O_{B}}(\left\{ (x,x) | x\in \mathbb{R}\right\}) = 1$. Such pairs of observables are typically called EPR-doubles: the outcome of measuring one predicts with certainty the outcome of measuring the other. These EPR-doubles form type I factors (isomorphic to von Neumann algebras $\textbf{B} (\mathcal{H}_{A}) \otimes \mathcal{I}$ and $\mathcal{I} \otimes \textbf{B} (\mathcal{H}_{B})$ respectively), so they have a discrete spectrum (their distributions relative to $\omega$ take discrete values only) (Arens and Varadarajan 2000, 647). Thus, this definition is not enough to characterize the original EPR state -- the state that Einstein was concerned with. In order to overcome this limitation, Reinhard F. Werner considered $\mathfrak{A}_{A} \otimes \mathcal{I}$ and $\mathcal{I} \otimes \mathfrak{A}_{B}$ as the mutually commuting subalgebras of $\mathfrak{A}_{A} \otimes \mathfrak{A}_{B}$, and took $\pi (\mathfrak{A}_{A} \otimes \mathcal{I})''$ as a self-adjoint unital subalgebra of $\textbf{B} (\mathcal{H}_{A}) \otimes \mathcal{I}$, and $\pi (\mathcal{I} \otimes \mathfrak{A}_{B})''$ as a self-adjoint unital subalgebra of $\mathcal{I} \otimes \textbf{B} (\mathcal{H}_{B})$, both closed in the weak operator topology. Then $\pi(O_{A}\otimes I) \in \pi (\mathfrak{A}_{A} \otimes \mathcal{I})''$ and $\pi( I \otimes O_{B}) \in \pi (\mathcal{I} \otimes \mathfrak{A}_{B})''$ are EPR-doubles and have continuous spectra, since $\pi (\mathfrak{A}_{A} \otimes \mathcal{I})''$ and $\pi (\mathcal{I} \otimes \mathfrak{A}_{B})''$ are type II$_{1}$ factors (Werner 1999). 

Now, consider the representations $(\pi, \mathcal{H}_{A} \otimes I)$ and $(\pi, I \otimes \mathcal{H}_{B})$ of $\mathfrak{A}_{A} \otimes \mathfrak{A}_{B}$. If $\omega$ is an original EPR state and $\pi$ is faithful, then there is a state $\tau$ in each of these representations such that for any element $O_{A} \otimes O_{B} \in \mathfrak{A}_{A} \otimes \mathfrak{A}_{B}$, we will have $\omega (O_{A} \otimes O_{B}) = \tau(\pi(O_{A} \otimes O_{B}))$ (Halvorson 2000, 327). This entails, on the same conditions, that for any two representations $(\pi_{1}, \mathcal{H}_{A} \otimes I)$ and $(\pi_{2}, \mathcal{H}_{A} \otimes I)$ of $\mathfrak{A}_{A} \otimes \mathcal{I}$, there are different states $\tau_{1}$ and $\tau_{2}$ in these representations, respectively, such that for two different elements $O^{1}_{A} \otimes I, O^{2}_{A} \otimes I \in \mathfrak{A}_{A} \otimes \mathcal{I}$, we have the corresponding restrictions of $\omega$, that is $\omega(O^{1}_{A} \otimes I) = \tau_{1}(\pi_{1} (O^{1}_{A} \otimes I))$ and $\omega(O^{2}_{A} \otimes I) =\tau_{2}(\pi_{2} (O^{2}_{A} \otimes I))$. All this suggests the following reformulation of the bottom half of Einstein's argument:

\bigskip

4$_{T}$. Quantum mechanics assigns different states, $\tau_{1}$ or $\tau_{2}$, to subsystem A, depending on which EPR-double, $\pi_{1} ( I \otimes O^{1}_{B})$ or $\pi_{2} ( I \otimes O^{2}_{B})$, one can choose to measure on B.

5$_{T}$. But if quantum mechanics is complete, $\tau_{1}$ and $\tau_{2}$ should be identical.

6$_{T}$. Thus, quantum mechanics is incomplete.

\bigskip

How should we understand premise 4$_{T}$? Within this framework, what makes $\tau_{1}$ and $\tau_{2}$ different states? As we have seen above, some commentators took Einstein's argument to be confused on this very point, for the reason that the only differences between such states allowed by standard quantum mechanics are ``trivial'' phase differences. But this is not what Einstein had in mind, as Howard already pointed out: ``Might there not be situations in which the differences between two $\psi$-functions (phase differences, for example) are inessential from the point of view of the system whose real state they aim to describe? Einstein's completeness
condition would, indeed, be too strong if it required that literally \textit{every} difference between $\psi$-functions
mirror a difference in the real state of the system in question; but such was not Einstein’s intention.'' (Howard 1985, 181) What are then the non-trivial differences between theoretical states that Einstein did have in mind? Here is Howard, again: ``The kind of difference with which Einstein was concerned is clear from his argument: [$\tau_{1}$] and [$\tau_{2}$] differ in the predictions they yield for the results of certain objective, local measurements on A. ... (For example, if [$\tau_{1}$] attributed a definite position to [A], but not a definite momentum, it would be incomplete in its description of [A]’s momentum; but, of course, Einstein's argument does not require any such reference to specific parameters or `elements of reality'.)'' (\textit{loc. cit.}, modified for uniform notation) The essential differences between  $\tau_{1}$ and $\tau_{2}$ concern their predictions of the measurement outcomes for A's observables. For instance, $\tau_{1}$ and $\tau_{2}$ are essentially different if one of them ``lives'' in $(\pi_{1}, \mathcal{H}_{A} \otimes I)$, say a position representation, and the other in $(\pi_{2}, \mathcal{H}_{A} \otimes I)$, say a momentum representation. More generally then, $\tau_{1}$ and $\tau_{2}$ are essentially different if $(\pi_{1}, \mathcal{H}_{A} \otimes I)$ and $(\pi_{2}, \mathcal{H}_{A} \otimes I)$ are unitarily inequivalent representations of $\mathfrak{A}_{A} \otimes \mathcal{I}$. Taking this into account, one obtains the following reconstruction of the bottom half of Einstein's argument:

\bigskip

4$^{T}$. Quantum mechanics allows unitarily inequivalent representations of $\mathfrak{A}_{A} \otimes \mathcal{I}$, depending on which EPR-double, $\pi_{1} ( I \otimes O^{1}_{B})$ or $\pi_{2} ( I \otimes O^{2}_{B})$, one can choose to measure on subsystem B.

5$^{T}$. But if quantum mechanics is complete, the representations $(\pi_{1}, \mathcal{H}_{A} \otimes I)$ and $(\pi_{2}, \mathcal{H}_{A} \otimes I)$ should be unitarily equivalent.

6$^{T}$. Thus, quantum mechanics is incomplete.

\bigskip

I believe that there is a good chance that this captures exactly what Einstein thought on the matter. But one immediate criticism against this reconstruction concerns the apparent conflict between premise 4$^{T}$ and the Stone-von Neumann theorem, conjectured by Marshall Stone (Stone 1930) and then proved by John von Neumann (von Neumann 1931), of which Einstein was no doubt fully aware by 1935. The theorem states that any irreducible faithful Hilbert space representation of the Weyl algebra generated by the canonical commutation relations describing a quantum mechanical system (or any system with a finite number of degrees of freedom) is uniquely determined up to a unitary transformation, and in fact unitarily equivalent to the Schrödinger representation.\footnote{This entails that all bounded operators are intertwined by an isometric isomorphism. For details, see Summers 2001,  Rosenberg 2004, Ruetsche 2011. The existence of this isomorphism has been considered sufficient to justify an interpretation of the Stone-von Neumann theorem as a categoricity result for quantum mechanics. For a critical discussion of such an interpretation, see Toader 2021.} 

The Stone-von Neumann theorem applies, of course, to EPR systems. But the conflict with my reconstruction is merely apparent, since the representations mentioned in premise 4$^{T}$ may be taken as non-regular and, therefore, not in the range of the Stone-von Neumann theorem.\footnote{There are other ways in which the conflict can be avoided, for example, by turning to phase spaces with nontrivial topologies, on which representations can be regular but unitarily inequivalent (Landsman 1990).} This reply, however, just seems to have opened the door to further criticism. For if Einstein's argument required that regularity be dropped, then one might seem justified to consider it muddled after all, for it is not at all clear what physical meaning can be given to non-regular representations. If, besides the Schrödinger representation, both position and momentum representations are taken to be physically significant despite their non-regularity, then unitary dynamics must be given up.\footnote{For details, see Feintzeig \textit{et al.} 2019, 126sq. This observation is essentially based on an unpublished result by David Malament, according to which if a free dynamics is assumed in the position representation, then exact localizabilty is violated.} But I think that this captures precisely the sense of incompatibility that Gömöri and Hofer-Szabó, as we have seen above, associate with the multiple assignment of $\psi$-functions to subsystem A. The dynamical incompatibility between (regular and) non-regular representations not only does not muddle Einstein's argument, but rather helps clarify his justification for rejecting completeness.

Now, if my algebraic reconstruction is correct, a formalization in a suitable metatheory would be required to fully support Howard's suggestion.\footnote{Given the metric completeness of the mathematical structures of quantum mechanics, this formalization would be impossible in classical first-order logic. For a survey of recent work in the formal semantics of C$^{*}$-algebras, see Lupini 2019.} But an informal version of the reconstructed argument should be enough here:

\bigskip

4$_{H}$. Quantum mechanics allows non-isomorphic models of its description of A, depending on which EPR-double one can choose to measure on subsystem B.

5$_{H}$. But if quantum mechanics is categorical, these models should be isomorphic.

6$_{H}$. Thus, quantum mechanics is not categorical.

\bigskip

As noted already, this version of the argument assumes that unitarily inequivalent representations (in premise 4$^{T}$) can be construed as non-isomorphic models of the quantum mechanical description of subsystem A (in premise 4$_{H}$). In other words, it assumes that unitary equivalence is a necessary, though not a sufficient, condition for categoricity. On this assumption, which I consider unproblematic, if the validity of the argument is to be preserved, premise 5$^{T}$ must be reformulated as premise 5$_{H}$. Einstein completeness should indeed be understood as categoricity. I think that this clarifies and supports, at least in part, Howard's ``outrageous'' suggestion. Einstein's no-go result is really that a local, separable, and categorical quantum mechanics cannot exist.%\footnote{This raises the question if a non-local quantum mechanics can be categorical. The answer will depend, of course, on what the models of the theory are taken to be, and on whether the relevant algebraic relations between them can be formally construed at least in principle in terms of a model-theoretical isomorphism.}

\section{Consequences}

Reconstructing Einstein's argument as above points to a series of general consequences often discussed, in other theoretical contexts, throughout the rich history of philosophical concerns with the categoricity of logic and mathematics. For instance, one often takes it to be the case that ``A non-categorical set of sentences ... does not give the impression of a closed and organic unity and does not seem to determine precisely the meaning of the concepts contained in it." (Tarski 1934, 311) The models of a categorical theory agree on all theorems or predictions that can be expressed in its language. But only a categorical theory determines the meaning of its concepts precisely. For instance, the meaning of arithmetical terms is precisely determined only if Peano arithmetic is categorical with respect to the isomorphism class of an omega sequence. The meaning of classical logical connectives is precisely determined only if classical logic is categorical with respect to the isomorphism class of a two-element Boolean algebra. This suggests that, quite similarly, the meaning of concepts like quantum state, observable, etc. is precisely determined only if quantum mechanics is categorical with respect to the relevant isomorphism class of models. Thus, Einstein's argument, if reconstructed as I proposed above, attributes to quantum mechanics a kind of semantic indeterminacy that may help explain the difference between EPR completeness and Einstein completeness as types of descriptive completeness. The question is what kind of descriptive failure is the incompleteness that Einstein attributed to quantum mechanics? As a non-categorical theory, quantum mechanics is descriptively incomplete, just not in the sense that it fails to describe all of the physical reality that it aims to describe, but rather in the sense that it describes more than just the physical reality that it aims to describe. This does not necessarily imply that the theory describes states that cannot exist; rather, it describes dynamically incompatible states, construed as (elements of) non-isomorphic models.

Why is the non-categoricity of quantum mechanics not considered a problem today? Why is it’s having fallen out of sight not itself considered as a serious problem? Why is categoricity not a primary goal of contemporary quantum physicists? These are important questions that I cannot address here. What I want to briefly emphasize before I conclude this paper are some implications I take my reconstruction of Einstein's argument to have for our understanding of his debate with Bohr.

Let us recall one important point in that debate, that is, that Bohr's reply to the EPR paper, as was well understood by Einstein, denies separability (Howard 2007). Thus, it also applies to Einstein's own argument (by rejecting the first half of premise 1). Einstein famously commented on this point as follows: ``By this way of looking at the matter it becomes evident that the paradox forces us to relinquish one of the following two assertions:
(1) the description by means of the $\psi$-function is \textit{complete}.
(2) the real states of spatially separated objects are independent of each other.'' (Einstein 1949, 682) Einstein thought that relinquishing assertion (2), i.e., denying that spatially separated objects have real states, would make physics impossible. But a second important point made by Bohr is that EPR-doubles are complementary (Bohr 1935). What does this point amount to, when considered in the algebraic framework described above? Bohr's complementarity has been rigorously interpreted in terms of the unitary inequivalence of non-regular Hilbert space representations (Halvorson 2004). Since Einstein's argument, on my reconstruction, points out that there are representations describing subsystem A of an EPR system that are unitarily inequivalent (premise 4$^{T}$), Bohr's reply can be understood as a rejection of premise 5$^{T}$: a complete quantum mechanics does not require unitary equivalence of representations. Furthermore, if it is correct to reformulate premise 5$^{T}$ as premise 5$_{H}$, then one can see Einstein and Bohr holding opposite positions in a spectrum of model-theoretical views regarding quantum mechanics: while Einstein would deplore its non-categoricity, Bohr would embrace it as a theoretical asset.

\section{Conclusion}

The reconstruction of Einstein's argument offered in this paper clarifies the assumptions behind Howard's suggestion to read Einstein completeness as categoricity. It also reinforces criticism of some of the misconstruals of Einstein completeness in the literature. Finally, it explains the sense in which Einstein completeness is different than EPR completeness as a type of descriptive completeness, and it sheds some light on the Bohr-Einstein controversy. A more thorough understanding of these issues would certainly benefit from a fully articulated formal semantics of quantum mechanics.

\section{Statement}

There are no conflicts of interest in connection to this paper.

\section{Acknowledgements}

I am grateful to Don Howard for making the suggestion discussed in this paper and for his now decades-long mentorship; to Guido Bacciagaluppi for reminding me of the suggestion, in Utrecht, in 2017; to Jos Uffink for drawing my attention to the interpretation of Einstein completeness as overcompleteness, in Salzburg, in 2019; and more recently, to Marton Gömöri, Gábor Hofer-Szabó, Sebastian Horvat, and Magdalena Zych for comments and discussion that helped improve the paper. Any remaining errors are my sole responsibility.  

\section{References}

Acuña, P. (2021) von Neumann’s Theorem Revisited, \textit{Foundations of Physics}, 51, 73

Arageorgis, A. (1995) \textit{Fields, particles, and curvature: Foundations and philosophical
aspects of quantum field theory in curved spacetime}, Ph.D. thesis, University of
Pittsburgh

Arens, R. and V. S. Varadarajan (2000) On the concept of Einstein–Podolsky–Rosen
states and their structure, \textit{Journal of Mathematical Physics}, 41, 638--651

Bohr, N. (1935) Can Quantum-Mechanical Description of Physical Reality Be Considered Complete? \textit{Physical Review}, 48, 696--702

Detlefsen, M. (2014) Completeness and the ends of axiomatization, \textit{Interpreting Gödel}, 59--77

Einstein A., B. Podolski, and N. Rosen (1935) Can Quantum-Mechanical Description of Physical Reality Be Considered Complete? \textit{Physical Review}, 47, 777--780

Einstein A. (1936) Physics and Reality, \textit{Ideas and opinions}, 1954, 290--323

Einstein A. (1949) Remarks Concerning the Essays Brought together in this Co-operative Volume, \textit{Albert Einstein: Philosopher-Scientist}, 665--688

Feintzeig, B., J.B. Le Manchak, S. Rosenstock, \& J.O. Weatherall (2019) Why Be regular?, part I, \textit{Studies in History and Philosophy of Modern Physics}, 65, 122--132

Fine, A. (1981) Einstein's Critique of Quantum Theory: The Roots and Significance of EPR, \textit{The Shaky Game}, 1986, 26--39

Gömöri, M. and G. Hofer-Szabó (2021) On the meaning of EPR’s Reality Criterion, \textit{Synthese} 199, 13441--13469.

Halvorson, H. (2000) The Einstein-Podolsky-Rosen State Maximally
Violates Bell's Inequalities, \textit{Letters in Mathematical Physics} 53, 321--329

Halvorson, H. (2004) Complementarity of representations in quantum mechanics, \textit{Studies in History and Philosophy of Modern Physics}, 35, 45--56

Howard, D. (1985) Einstein on Locality and Separability, \textit{Studies in History and Philosophy of Science}, 16, 171--201 

Howard, D. (1990) `Nicht Sein Kann Was Nicht Sein Darf', Or the Prehistory of the EPR, 1909-1935: Einstein's Early Worries About the Quantum Mechanics of Composite Systems, \textit{Sixty-Two Years of Uncertainty: Historical, Philosophical, and Physical Inquiries into the Foundations of Quantum Mechanics}, ed. by A. I. Miller, Plenum Press, 61--111

Howard, D. (1992) Einstein and \textit{Eindeutigkeit}: A Neglected Theme in the Philosophical Background to General Relativity, \textit{Studies in the History of General Relativity}, ed. by J. Eisenstaedt and A. J. Kox, Birkhäuser, 154--243

Howard, D. (2007) Revisiting the Einstein-Bohr Dialogue, \textit{Iyyun: The Jerusalem Philosophical Quarterly}, 56, 57--90

Hughes, R. I. G. (1989) \textit{The Structure and Interpretation of Quantum Mechanics}, Harvard University Press

Landsman, N. P. (1990) C$^{*}$-algebraic quantization and the origin of topological quantum effects, \textit{Letters in Mathematical Physics}, 20, 11--18

Landsman, N. P. (2006) When champions meet: Rethinking the Bohr–Einstein debate, \textit{Studies in History and Philosophy of
Modern Physics}, 37, 212--242 

Lehner, C. (2014) Einstein's Realism and His Critique of Quantum Mechanics, \textit{The Cambridge Companion to Einstein}, CUP, 306--353

Lupini, M. (2019) An Invitation to Model Theory and C$^{*}$-Algebras, \textit{The Bulletin of Symbolic Logic}, 25, 34--100  

McGrath, J. H. (1978) A Formal Statement of
the Einstein-Podolsky-Rosen Argument, \textit{International Journal of Theoretical Physics}, 17, 557--571

Murgueitio Ram\'irez, S. (2020) Separating Einstein's separability, \textit{Studies in History and Philosophy of Modern Physics}, 72, 138--149

Norsen, T. (2017) \textit{Foundations of Quantum
Mechanics}, Springer 

Rosenberg, J. (2004) A selective history of the Stone–von Neumann Theorem, \textit{Operator Algebras, Quantization, and Noncommutative Geometry}, Providence, RI, American Mathematical Society, 331--354

Ruetsche, L. (2011) \textit{Interpreting Quantum Theories. The Art of the Possible}, Oxford University Press.

Sauer, T. (2007) An Einstein manuscript on the EPR paradox for spin observables, \textit{Studies in History and Philosophy of
Modern Physics} 38, 879–-887

Stone, M. H. (1930) Linear Transformations in Hilbert Space, \textit{Proceedings of the National Academy of Sciences of the USA}, 16, 172--175

Summers, S. J. (2001) On the Stone-von Neumann Uniqueness Theorem and Its Ramifications, \textit{John von Neumann and the Foundations of Quantum Physics}, ed. by M. R\'edei and M. St\"oltzner, Springer, 135--152

Tarski, A. (1934) Some methodological investigations on the definability of concepts, \textit{Logic, Semantics, Metamathematics}, Hackett, 298--319

Toader, I. D. (2021) On the Categoricity of Quantum Mechanics, \textit{European Journal for Philosophy of Science}, 11, 17

Uffink, J. (2020) Schrödinger’s reaction to the EPR paper. \textit{Quantum,
Probability, Logic}, ed. by M. Hemmo and O. Shenker, Springer, 545--566

von Neumann, J. (1931) Die Eindeutigkeit der Schr\"{o}dingerschen Operatoren, \textit{Mathematische Annalen}, 104, 570--578

Werner, R. F. (1999) EPR states for von Neumann algebras, \textit{arXiv:quant-ph/9910077}

Weatherall, J. O. (2019) Part 1: Theoretical equivalence in physics, \textit{Philosophy Compass}, 10.1111/phc3.12592

\end{document}